\documentclass{tCTM2e}
\usepackage{cite,upmath}
\usepackage{amsmath,amssymb}

\newcommand{\dd}{\ensuremath{\mathrm{d}}}
\newcommand{\DD}{\ensuremath{\mathrm{D}}}
\newcommand{\vect}[1]{\ensuremath{{\boldsymbol #1}}}
\newcommand{\tens}[1]{\ensuremath{{\boldsymbol{\mathsf{#1}}}}}

\title{Level set simulations of turbulent thermonuclear
  deflagration in degenerate carbon and oxygen}

\author{W.~Schmidt, Lehrstuhl f\"{u}r Astronomie, Universit\"{a}t
  W\"{u}rzburg, Am Hubland, D-97074 W\"{u}rzburg, Germany\\
	W.~Hillebrandt, Max-Planck-Institut f\"{u}r Astrophysik,
  Karl-Schwarzschild-Str.\ 1, D-85741 Garching, Germany\\
	J.~C.~Niemeyer, Lehrstuhl f\"{u}r Astronomie, Universit\"{a}t
  W\"{u}rzburg, Am Hubland, D-97074 W\"{u}rzburg, Germany}

\begin{document}

\maketitle

\begin{abstract}
  We study the dynamics of thermonuclear flames propagating in fuel
  stirred by stochastic forcing. The fuel consists of carbon and
  oxygen in a state which is encountered in white dwarfs close to the
  Chandrasekhar limit. The level set method is applied to represent
  the flame fronts numerically.  The computational domain for the
  numerical simulations is cubic, and periodic boundary conditions are
  imposed. The goal is the development of a suitable flame speed model
  for the small-scale dynamics of turbulent deflagration in
  thermonuclear supernovae.  Because the burning process in a
  supernova explosion is transient and spatially inhomogeneous, the
  localised determination of subgrid scale closure parameters is
  essential. We formulate a semi-localised model based on the
  dynamical equation for the subgrid scale turbulence energy
  $k_{\mathrm{sgs}}$. The turbulent flame speed $s_{\mathrm{t}}$ is of
  the order $\sqrt{2k_{\mathrm{sgs}}}$. In particular, the subgrid
  scale model features a dynamic procedure for the calculation of the
  turbulent energy transfer from resolved toward subgrid scales, which
  has been successfully applied to combustion problems in
  engineering. The options of either including or suppressing inverse
  energy transfer in the turbulence production term are compared.  In
  combination with the piece-wise parabolic method for the
  hydrodynamics, our results favour the latter option. Moreover,
  different choices for the constant of proportionality in the
  asymptotic flame speed relation,
  $s_{\mathrm{t}}\propto\sqrt{2k_{\mathrm{sgs}}}$, are investigated.
\end{abstract}


\section{Introduction}

A certain kind of stellar explosion, known as \emph{type Ia
supernovae} among astronomers, is currently explained by the
thermonuclear explosion of an electron-degenerate stellar remnant
\cite{HoyFow60}.  Such an object, which is called a white dwarf,
emanates from the burn-out of stars comparable in mass to our Sun and
is mainly composed of carbon and oxygen. If the white dwarf has a
companion star in close orbit, it can grow by accreting material from
the companion. Under certain conditions, the white dwarf's mass will
steadily increase and finally approach the \emph{Chandrasekhar limit},
which is the maximal mass that can be supported by the degeneracy
pressure of electrons \cite{ShaTeuk}. As the temperature and density
are increasing, thermonuclear burning of carbon and oxygen gradually
sets in.  Close to the Chandrasekhar mass, the conditions in the core
of the white dwarf eventually pass a critical threshold
\cite{WoosWun04}. At this point, the rate of thermonuclear reactions
rises rapidly, and a runaway is initiated, which incinerates and
disrupts the whole star within a few seconds. The total energy release
is of the order $10^{51}\,\mathrm{erg}$ \cite{HilleNie00}.

The thermonuclear combustion of degenerate carbon and oxygen of
density in the range $\sim 10^{7}\ldots 10^{9}\,\mathrm{g\,cm^{-3}}$
proceeds in the form of a deflagration
\cite{NomThiel84,TimWoos92}. Since the nuclear ash produced by the
burning process has less specific weight than the surrounding
unprocessed material, it becomes Rayleigh-Taylor unstable. Since
turbulence is subsequently produced, the flames get corrugated and
folded \cite{Khok95,BellDay04}. In consequence, there is a positive
feedback mechanism of turbulence enhancing the burning and, according
to state-of-the-art numerical simulations, eventually results in an
explosion \cite{NieHille95a,ReinHille02b,GamKhok03}. Although it
cannot be ruled out that a transition from the deflagration to a
detonation might set in at some stage \cite{Khok91a,GamKhok04},
turbulent deflagration plays a crucial role in the theoretical
modelling of thermonuclear supernovae in any case.

The subject of this article is the dynamics of flame fronts on length
scales much smaller than the size of a Chandrasekhar-mass white
dwarf. To that end, an artificial scenario was set up. A turbulent
flow is produced by means of stochastic stirring in a cubic domain
subject to periodic boundary conditions \cite{EswaPope88,Schmidt04}.
Thermonuclear burning is ignited in small spherical regions and
subsequently evolved by means of the \emph{level set method}
\cite{OshSeth88,ReinHille99a}.  The complicated network of
thermonuclear reactions encountered in a type Ia supernova is
substituted by the effective fusion of equal mass fractions of
${^{12}}$C and ${^{16}}$O to ${^{56}}$Ni and ${^{4}}$He as
representative reaction \cite{SteinMuel92}.  The equation of state is
dominated by the degenerate gas of relativistic free electrons. Thus,
the approximate relation $P\propto\rho^{4/3}$ applies, while the
temperature has virtually no influence on the pressure. This is
actually the reason for the runaway, because the negative feedback
between heating and expansion in non-degenerate matter is absent.  The
exact equation of state has no analytic solution and must be
integrated numerically. Moreover, contributions from nuclei, photons
and pair electron-positron pair creation at temperatures of the order
$10^{10}\,\mathrm{K}$ are taken into account (section 3.2 of
\cite{Rein01}).  The fluid dynamics is treated by means of the
piece-wise parabolic method (PPM) within the framework of the Euler
equations \cite{CoWood84}.

In the corrugated flamelet regime of combustion, the flame propagation
is affected by turbulence on length scales ranging from the Gibson
length up to the integral length scale \cite{NieKer97}. In general,
only the the largest length scales can be resolved in a numerical
simulation. In order to account for the wrinkling of the flame surface
on length scales smaller than the cutoff scale of a simulation, an
effective propagation speed, the so-called turbulent flame speed, must
be calculated.  This involves a subgrid scale (SGS) model for the
local budget of turbulence energy contained in numerically unresolved
modes.  In this article, we present a numerical study in which different
variants of the SGS turbulence energy model are compared (section 4.3
in \cite{Sagaut}). In particular, we adopted a dynamical procedure for
the computation of SGS closure parameters. This procedure was proposed
by Kim and Menon for the application in LES of gas turbine combustor
flows \cite{KimMen99}.

\section{The physics of turbulent thermonuclear deflagration}
\label{sc:lam_burn}

The mechanism of deflagration is based on thermal conduction, as
opposed to a detonation, which proceeds via shock compression.
Unburned material (fuel) is heated in the vicinity of the reaction
zone and thereby gets ignited.  Once heat generation is balanced by
diffusion, the burning zone is propagating at a steady subsonic speed,
and pressure equilibrium is maintained across the reaction zone.
Basically, this characterises what is commonly known as a
\emph{flame}. For chemical combustion, a distinction is made between
\emph{premixed} and \emph{diffusive} flames.  Thermonuclear flames are
trivially premixed, because no additional agent, like oxygen in most
chemical burning processes, is required. The local propagation speed
of the flame, which is solely determined by microscopic properties, is
called the \emph{laminar burning speed}.  The notion of a flame
applies, if fluid motions do not significantly disturb the burning
process within the reaction zone, i.\ e., the characteristic time
scale of burning is much smaller than the kinetic time scale of
velocity fluctuations on length scales comparable to the flame
thickness $\delta_{\mathrm{F}}$. Equivalently, $\delta_{\mathrm{F}}\ll
l_{\mathrm{G}}$, where the Gibson length scale $l_{\mathrm{G}}$ is the
smallest length scale on which the burning process is affected by
fluid motion \cite{NieKer97}. The condition $\delta_{\mathrm{F}}\ll
l_{\mathrm{G}}$ thus specifies the \emph{flamelet regime} of
combustion, which is reviewed in this section. For the thermonuclear
combustion in C+O white dwarfs, it appears that the flamelet
description is valid for $\rho\gtrsim 3\cdot 10^{7}\,\mathrm{g\,cm^{-3}}$
\cite{NieWoos97}. In thermonuclear supernovae, most of the burning
takes place at significantly higher densities.

\subsection{Laminar burning}

The width of the reaction zone, $\delta_{\mathrm{F}}$, is determined
by the equilibrium between energy generation due to nuclear reactions
and the rate of diffusion caused by thermal conduction (\S~128 in
\cite{LanLifVI}). The balance between these processes can be expressed
in terms of their characteristic time scales, $\tau_{\mathrm{burn}}$
and $\tau_{\mathrm{cond}}$. The former is given by
$\tau_{\mathrm{burn}}\sim
\rho\varepsilon_{\mathrm{nuc}}/B$, where
$\varepsilon_{\mathrm{nuc}}$ is the energy generated by the fusion of
a unit mass of nuclear fuel, and $B$ is the rate of
energy release per unit volume.  The time scale of conduction, on the
other hand, can be expressed as $\tau_{\mathrm{cond}}\sim
\delta_{\mathrm{F}}^{2}/l_{\mathrm{e}}c$, where $l_{\mathrm{e}}$ is
the mean free path of the electrons, which contribute the major part
of the thermal conductivity, and $c$ is the speed of sound. Setting
these two time scales equal, one finds that the \emph{flame thickness}
$\delta_{\mathrm{F}}$ is approximately given by
\begin{equation}
  \label{eq:flame_width}
  \delta_{\mathrm{F}} \sim
  \sqrt{\frac{\rho\varepsilon_{\mathrm{nuc}}l_{\mathrm{e}}c}{B}}.
\end{equation}
Defining the \emph{laminar flame speed} by
$s_{\mathrm{lam}}=\delta_{\mathrm{F}}/\tau_{\mathrm{burn}}$, we have
\begin{equation}
  s_{\mathrm{lam}} \sim 
  \sqrt{\frac{l_{\mathrm{e}} c B}{\rho\varepsilon_{\mathrm{nuc}}}}.
\end{equation}
The specific energy release for the fusion of ${^{12}}$C and
${^{16}}$O to ${^{56}}$Ni is $\varepsilon_{\mathrm{nuc}}\approx 7\cdot
10^{17}\mathrm{erg\,g^{-1}}$ \cite{SteinMuel92}. The flame speed
$s_{\mathrm{lam}}$ for the thermonuclear combustion of degenerate
carbon and oxygen was computed numerically for a wide range of mass
densities and nuclear compositions by Timmes and Woosley
\cite{TimWoos92}. For example, $s_{\mathrm{lam}}\approx
3.6\cdot 10^{6}\,\mathrm{cm\,s^{-1}}$ and $\delta_{\mathrm{F}}\approx 2.9\cdot
10^{-4}\,\mathrm{cm}$ for equal mass fractions of carbon and oxygen at
a density $10^{9}\,\mathrm{g\,cm^{-3}}$.

\subsection{Turbulent burning}
\label{sc:turb_burn}

So far, we have only been concerned with the microphysics of
thermonuclear deflagration. Let us now consider the combustion of
C$+$O fuel in a state of turbulent motion. For brevity, we shall
assume the case of steady isotropic turbulence, i.e.\ a statistically
self-similar hierarchy of vortices or eddies. Each vortex of size $l$ has a
characteristic velocity $v'(l)$ and an associated turn-over
time $\tau_{\mathrm{eddy}}(l)=l/v'(l)$. If $v'(l)$ is
small compared to the laminar flame speed $s_{\mathrm{lam}}$, then the
flame front will propagate through a region of size $l$ in a time much
faster than the turn-over time $\tau_{\mathrm{eddy}}(l)$. Hence, the
turbulent flow appears to be more or less ``frozen'' with respect to
the burning process on these scales.  For $v'(l)\gg
s_{\mathrm{lam}}$, on the other hand, the front is significantly
distorted while it is crossing a vortex of diameter $l$.  Hence, there
is a threshold scale on which burning decouples from turbulence. This
is the \emph{Gibson length} $l_{\mathrm{G}}$, which is defined by \cite{NieWoos97}
\begin{equation}
  v'(l_{\mathrm{G}}) = s_{\mathrm{lam}}.
\end{equation}
At length scales $l\gg l_{\mathrm{G}}$, turbulence corrugates the
flame and thereby increases its surface area. Consequently, turbulence
enhances the burning process and the release of heat is growing.  This
can be accounted for by introducing a \emph{turbulent propagation
speed} $s_{\mathrm{t}}(l)$.  In other words, averaging over regions of
size $l$, the flame front propagates with an effective speed
$s_{\mathrm{t}}(l)$ greater than the laminar burning speed
$s_{\mathrm{lam}}$, which specifies the local speed of any portion of
the flame smaller than $l_{\mathrm{G}}$. 

For $l\gg l_{\mathrm{G}}$, $s_{\mathrm{t}}(l)$ becomes asymptotically
independent of $s_{\mathrm{lam}}$.  The fundamental hypothesis applied
in this article is that $s_{\mathrm{t}}(l)$ is then given by the
magnitude of the turbulent velocity fluctuations $v'(l)$. In the
framework of the phenomenological Kolmogorov theory of isotropic
turbulence \cite{Kolmog41}, this velocity obeys the scaling law
$v'(l)\propto l^{1/3}$ in the inertial subrange, i.e.\ in the range
of length scales which are neither affected by the viscosity of the
fluid nor large-scale energy injection. Therefore,
\begin{equation}
  \label{eq:gibson_scaling}
  s_{\mathrm{t}}(l)\sim v'(l)\propto l^{1/3}.
\end{equation}
The relation $s_{\mathrm{t}}(l)\sim v'(l)$ was first proposed by
Damk\"{o}hler, who studied Bunsen cones in the laboratory
\cite{Damk40}.  Further validation of this conjecture came from
numerical studies \cite{Kerst88}. A motivation based on a theoretical
analysis in the framework of the level set prescription was given
by Peters \cite{Peters99}. 

\section{The numerical modelling of turbulent flame propagation}
\label{sc:num_model}

In early studies of thermonuclear deflagration
\cite{Khok95,NieHille95a}, a \emph{reactive-diffusive} flame model
with artifical diffusion and reaction rates was applied.  In this
approach, the thickness of the flame is artifically increased over
several grid cells and the propagation speed is adjusted to a
prescribed value. On the other hand, 
the \emph{level set method} proposed by Osher and Sethian
\cite{OshSeth88,Sethian} is a front tracking method which describes
the interface separating ash from fuel as a genuine discontinuity.
The interface is numerically represented by the set
of all points for which a suitably chosen \emph{distance function}
vanishes, i.e.\ the zero level set.  This is a sensible approximation
if the physical flame thickness is very small compared to the Gibson
scale.  For the simulation of thermonuclear combustion in type Ia
supernovae, the level set method was implemented by Reinecke
\cite{ReinHille99a,Rein01}.

\subsection{The level set method}

Let $G(\vect{x},t)$ be a signed distance function with the
property $|\vect{\nabla} G|=1$. The absolute value
$|G(\vect{x},t)|$ is equal to the minimal distance of the point
$\vect{x}$ from the flame front at time $t$. The front itself is
given by the constraint $G(\vect{x},t)=0$, i.e.\ it is
represented by the zero level set
$\Gamma(t)=\{\vect{x}|G(\vect{x},t)=0\}$.  With the sign
convention $G(\vect{x},t)>0$ in regions containing burned
material, the unit normal vector pointing towards unburned material is
given by $\vect{n}=-\vect{\nabla}
G/|\vect{\nabla}G|$.  The time evolution of the front
$\Gamma(t)$ is implicitly determined by the total time derivative of
$G(\vect{x}_{\Gamma}(t),t)=0$. For a certain point at the front,
$\vect{x}_{\Gamma}(t)\in\Gamma(t)$, we have
\begin{equation}
  \label{eq:lset_loc}
  \frac{\dd}{\dd t}G(\vect{x}_{\Gamma}(t),t) = 
  \frac{\partial G}{\partial t} +
  \dot{\vect{x}}_{\Gamma}\cdot\vect{\nabla} G = 0.
\end{equation}
The speed function $\dot{\vect{x}}_{\Gamma}$ is given by the sum of
two contributions.  Firstly, the advection speed normal to the flame
front, $\vect{v}_{\mathrm{u}}\cdot\vect{n}$, where
$\vect{v}_{\mathrm{u}}$ is the velocity of the fuel immediately ahead
of the front in an Eulerian frame of reference. And, secondly, the
intrinsic propagation speed $s$ of the flame front relative the fuel.

The local equation~(\ref{eq:lset_loc}) can be formulated globally as well,
without constraining the position $\vect{x}$ to the flame surface.
Substituting the definition of the normal vector $\vect{n}$ and
expressing the speed function in the form
$\vect{v}_{\mathrm{u}}\cdot\vect{n}+s$, the evolution equation for the
level set function at any point in space becomes
\begin{equation}
  \label{eq:lset}
  \frac{\partial G(x,t)}{\partial t} = 
  [\vect{v}_{\mathrm{u}}(x,t)+s(x,t)\vect{n}(x,t)]|\vect{\nabla} G(x,t)|.
\end{equation}
The advection part on the right-hand side can be treated with a
finite-volume scheme, for instance, the PPM.  The intrinsic front
propagation is usually calculated by means of an entropy-satisfying
upwind scheme.  In general, non-planar fronts will develop sharp
corners and the corresponding level set must be a weak solution:
Information about the initial conditions is lost, once a cusp has
formed, and the subsequent evolution is irreversible. The
corresponding entropy condition can be formulated in the following
way: Once a certain fluid element is burned, it remains burned
thereafter. In fact, this implies the equivalent Huyghen's principle
in optics for the propagation of the front over an infinitesimal
interval of time (section~5 in \cite{Sethian}). Finally, in order to
preserve the property $|\vect{\nabla} G|=1$, the updated
distance function has to be corrected after each time step. In the
implementation of Reinecke, this is achieved by means of
\emph{re-initialisation} \cite{ReinHille99a}.

For the complete implementation of the level set technique, both the
burned and the unburned state in an intersected numerical cell must
be reconstructed from the jump conditions across the front. Assuming that
there is a volume fraction of unburned material $\alpha$, conservation of
momentum imposes the constraint
\begin{equation}
  \rho\vect{v} = 
  \alpha\rho_{\mathrm{u}}\vect{v}_{\mathrm{u}} +
  (1-\alpha)\rho_{\mathrm{b}}\vect{v}_{\mathrm{b}},
\end{equation}
given the finite-volume averages $\rho$ and $\vect{v}$. The
volume fraction $\alpha$ can be calculated by linearly interpolating
the discrete numerical values of the distance function $G$.
Supplementing the momentum equation with the Rayleigh criterion, the
Hugoniot jump condition and the continuity constraint for the
tangential velocity components, a non-linear system of equations is
obtained. The solution yields $\vect{v}_{\mathrm{u}}$,
$\vect{v}_{\mathrm{b}}$ and the corresponding state variables
(cf.~\cite{RoepNie03}). This procedure of \emph{in-cell
reconstruction} was indeed successfully implemented for chemical
combustion problems \cite{SchmKlein03}.  However, the deviation of the
interpolated front element from the exact smooth solution can
introduce significant errors in the reconstructed states. In
particular, it is sometimes impossible to reconstruct physically sound
states for degenerate matter, because of the stiffness of the equation
of state. Moreover, one faces topological ambiguities for certain
configurations. A pragmatic method is to average over all possible
values, whenever one of these rare cases is encountered
\cite{ReinHille99a}.  Although R\"{o}pke \textit{et al.} have recently
succeeded with the implementation of in-cell reconstruction for the
problem of thermonuclear flame propagation in two dimensions
\cite{RoepNie03}, generalising the algorithm to three dimensions would
be much more challenging.

The difficulties outlined above are avoided with the so-called
\emph{passive} implementation, where the difference between burned and
unburned states is neglected and the advection speed is set equal to
$\vect{v}\cdot\vect{n}$. The discrete values of the velocity and state
variables are then interpreted as cell-centred averages. This is a
fair approximation in the limit of moderate density jumps between fuel
and ash.  A caveat of using the passive implementation for simulations
of burning at low density is the generation of numerical artifacts.
Fortunately, these problems are mainly encountered for densities
significantly less than about $10^{8}\,\mathrm{g\,cm^{-3}}$. Apart
from the systematic errors introduced by the averaged density and
advection velocity, the burning zone is not strictly represented by
the zero level set. Actually, there is a mixed phase between the
regions containing pure ash and fuel, respectively.  The width of the
diffusive smearing of the flame is typically a few cells, which is
still less than for the reaction-diffusion method. It was
demonstrated by numerous applications in simple test problems as well
as large-scale simulations of thermonuclear supernovae that the
passive implementation gives a satisfactory representation of the flame
fronts at high density and is robust even in three dimensions
\cite{ReinHille99b,ReinHille02b}.  For this reason, we used the
passive implementation for the simulations presented in this article.

\subsection{The turbulent flame speed relation}

On length scales larger than the Gibson length $l_{\mathrm{G}}$, flames are
predominantly shaped by turbulence.  If $l_{\mathrm{G}}$ is
significantly smaller than the numerical resolution $\Delta$, the
computed flame front appears inevitably smoother than its physical
counterpart. Consequently, the predicted burning rate would be
underestimated, if just the laminar burning speed was substituted for
the intrinsic propagation speed $s$ in equation~(\ref{eq:lset}).  This
is where the notion of the turbulent flame speed comes in.  We propose
that $s_{\mathrm{t}}(\Delta)$ is given by the local magnitude
of unresolved turbulence velocity fluctuations $v'(\Delta)$
\cite{NieHille95a}, albeit the turbulent flame speed, in a strict
sense, is an ensemble average. Since the propagation speed cannot be
less than the laminar flame speed, the simplest relation with
correct asymptotic behaviour is
\begin{equation}
  \label{eq:sgs_flame_speed_max}
  s_{\mathrm{t}}(\Delta)=
  \max(s_{\mathrm{lam}},\sqrt{2C_{\mathrm{t}}k_{\mathrm{sgs}}})=
  \max(s_{\mathrm{lam}},\sqrt{C_{\mathrm{t}}}q_{\mathrm{sgs}}).
\end{equation}
The quantity $k_{\mathrm{sgs}}=\frac{1}{2}q_{\mathrm{sgs}}^{2}$ is the
subgrid scale turbulence energy and $q_{\mathrm{sgs}}\sim v'(\Delta)$
the corresponding speed. An exact definition will be given in next
section.  For brevity, it is understood that $s_{\mathrm{t}}$ denotes
the turbulent flame speed at the numerical cutoff $\Delta$ in the
following.

For $\Delta\gg l_{\mathrm{G}}$, we have the asymptotic relation
$s_{\mathrm{t}}\approx \sqrt{C_{\mathrm{t}}}q_{\mathrm{sgs}}$ in the
limit of fully developed turbulence. Consequently, the turbulent flame
speed becomes independent of the laminar flame speed, and the
parameter $\sqrt{C_{\mathrm{t}}}$ determines the asymptotic scaling of
the turbulent flame speed. However, it is not quite clear whether the
constant of proportionality in the relation between
$s_{\mathrm{t}}(\Delta)$ and $q_{\mathrm{sgs}}$ is just unity or a
different value. Empirically, it appears that
$s_{\mathrm{t}}(\Delta)=2v'(\Delta)$, where
$v'(\Delta)=q_{\mathrm{sgs}}/\sqrt{3}$ \cite{Peters99}. Thus,
$C_{\mathrm{t}}=4/3$ in agreement with a constant of proportionality
close to unity.  Another shortcoming of the maximum
relation~(\ref{eq:sgs_flame_speed_max}) is that it gives a good
approximation to the turbulent flame speed in the laminar and the
fully turbulent regime, respectively, but not for the transition in
between. If $q_{\mathrm{sgs}}\sim s_{\mathrm{lam}}$, the relation
between turbulent flame speed and turbulence velocity might very well
be different.  For example, Im \textit{et al.} mention a quadratic
dependence on the turbulence velocity in the case of weak turbulence
\cite{ImLund97}. On the other hand, R\"{o}pke \textit{et al.} report a
linear relation even for turbulence velocities which are only
marginally larger than the laminar flame speed in two-dimensional
numerical simulations \cite{RoepHille04}. However, this result is
possibly unsubstantial for the three-dimensional case. Apart from
that, the transition from laminar to turbulent burning progresses
rather quickly, and a correct description in the intermediate phase is
therefore not overly important.

A different turbulent flame speed model was motivated theoretically by
Pocheau \cite{Poch94}:
\begin{equation}
  \label{eq:sgs_flame_speed_pocheau}
  \frac{s_{\mathrm{t}}}{s_{\mathrm{lam}}} = 
  \left[1 + C_{\mathrm{t}}
            \left(\frac{q_{\mathrm{sgs}}}{s_{\mathrm{lam}}}\right)^{n}
  \right]^{1/n}.
\end{equation}
In the scale-invariant regime, with $q_{\mathrm{sgs}}\gg
s_{\mathrm{lam}}$, the asymptotic form $s_{\mathrm{t}}\simeq
C_{\mathrm{t}}^{1/n}q_{\mathrm{sgs}}$ is obtained.  Kim \textit{et
al.}, chose $n=2$ and $C_{\mathrm{t}}\approx 20/3$ for LES of gas
turbine combustor flows \cite{KimMen99}. This value was inferred from
several laboratory experiments with hydrocarbon/air flames.
However, the data points cover values of the turbulent flame speed of
the same order as magnitude as the laminar burning speed only. For
this reason, one must be careful with any extrapolation to the fully
turbulent regime, in which $s_{\mathrm{t}}\gg s_{\mathrm{lam}}$.  If
$q_{\mathrm{sgs}}\ll s_{\mathrm{lam}}$, Taylor series expansion of the
right-hand side of equation~(\ref{eq:sgs_flame_speed_pocheau}) yields
the \emph{Calvin-Williams} relation,
\begin{equation}
  \frac{s_{\mathrm{t}}}{s_{\mathrm{lam}}} = 
  1 + C_{\mathrm{t}}
      \left(\frac{q_{\mathrm{sgs}}}{s_{\mathrm{lam}}}\right)^{2},
\end{equation}
which is consistent with the numerical results of Im \textit{et al.} 
\cite{ImLund97}.  

Apart from calculating the turbulent flame speed, secondary SGS
effects can be included in the dynamical equation~(\ref{eq:lset}) for
the level set function \cite{KimMen99,Peters99}. This was indeed numerically
investigated by Im \textit{et al.}  \cite{ImLund97}. In particular,
they suggested a procedure for the computation of $C_{\mathrm{t}}$ in
the fashion of the localised closure for the production parameter
$C_{\nu}$, which will be discussed in
section~\ref{sc:sgs_locl}. Whether this is advisable in combination
with the passive implementation, where numerical artifacts in the
shape of the resolved level set might produce significant spurious
contributions, is questionable. For this reason, it has not been
attempted.  In addition, there is a SGS transport term for the level
set, which is of the form
$\partial_{k}(\langle\overset{\infty}{v_{k}}\overset{\infty}{G}\rangle_{\mathrm{eff}}
-v_{k}G)$ and effectively introduces \emph{diffusion} of the level set
due to SGS turbulence. However, Kim \textit{et al.} argued that the
contributions arising thereof are not particularly important and, in
fact, cannot be determined within the available framework of SGS
modelling \cite{KimMen99}.

\section{The subgrid scale model}
\label{sc:sgs}

For the determination of the turbulent flame speed according to
equation~(\ref{eq:sgs_flame_speed_max})
or~(\ref{eq:sgs_flame_speed_pocheau}), the kinetic energy
$k_{\mathrm{sgs}}$ of unresolved vortices has to be computed.  A
dynamical equation for $k_{\mathrm{sgs}}$ is obtained through
decomposition of the conservation law for kinetic energy.  The
procedure of decomposing is conceptually based on the notion of
filtered quantities. In general, a filter is a convolution operator,
which smoothes out fluctuations on spatial scales smaller than the
characteristic length of the filter.  If a certain numerical solution
of the hydrodynamical equations is computed by means of a
finite-volume scheme, say, the PPM, then one can associate this
solution with the smoothed velocity field $\vect{v}(\vect{x},t)$,
which is obtained by mass-weighted or Favre filtering of the exact
realisation of the flow:
\begin{equation}
    \vect{v}(\vect{x},t) = 
    \frac{\langle\overset{\infty}{\rho}(\vect{x},t)
                 \overset{\infty}{\vect{v}}(\vect{x},t)\rangle_{\mathrm{eff}}}
         {\rho(\vect{x},t)},
\end{equation}
where
$\rho(\vect{x},t)=\langle\overset{\infty}{\rho}(\vect{x},t)\rangle_{\mathrm{eff}}$
is the smoothed mass density. The underlying hypothesis is that, if
the physical flow $\overset{\infty}{\vect{v}}(\vect{x},t)$ were known,
there exists a filter $\langle\ \rangle_{\mathrm{eff}}$ with suitable
properties such that the Favre-filtered velocity field $
\vect{v}(\vect{x},t)$ would reproduce the numerically computed
velocity field. Corresponding to the smoothed and the fluctuating
components of $\overset{\infty}{\vect{v}}(\vect{x},t)$, respectively,
one can distinguish the resolved part,
$k_{\mathrm{res}}=\frac{1}{2}|\vect{v}|^{2}$, and the subgrid
scale part $k_{\mathrm{sgs}}$ of the specific kinetic energy. In the
following, a formal decomposition of the kinetic energy is devised and
the dynamical equation for $k_{\mathrm{sgs}}$ is formulated. The
non-linearity of the conservation laws necessitates closure relations
for several terms in the decomposed equations. SGS closures and the
calculation of associated parameters are discussed in the remainder of
this section.

\subsection{The subgrid scale turbulence energy model}

In \emph{Germano's consistent decomposition}, the SGS turbulence
energy is simply defined by the difference between smoothed and
resolved kinetic energy \cite{Sagaut,Germano92}. This decomposition is
equivalent to setting $k_{\mathrm{sgs}}=-\frac{1}{2}\tau_{ii}$, where
the SGS turbulence stress tensor $\tau_{ik}$ is defined by
\begin{equation}
  \tau_{ik} \equiv 
  \tau(\overset{\infty}{v_{i}},\overset{\infty}{v_{k}}) =
  -\langle\overset{\infty}{\rho}\overset{\infty}{v_{i}}
  \overset{\infty}{v_{k}}\rangle_{\mathrm{eff}} +
  \rho v_{i} v_{k}.
\end{equation}
Hence, the SGS turbulence energy is given by
\begin{equation}
  k_{\mathrm{sgs}}=\frac{1}{2}\left[\frac{1}{\rho}
    \langle\overset{\infty}{\rho}|\overset{\infty}{\vect{v}}|^{2}\rangle_{\mathrm{eff}} -
    |\vect{v}|^{2}\right].
\end{equation}
We prefer the Germano decomposition, because it yields the
conceptually most transparent definition of the SGS turbulence energy
and avoids formal difficulties associated with SGS
closures. Deviations of the turbulent flame speed in alternative
decompositions correspond to higher-order terms which result from
secondary filtering of filtered quantities\cite{Sagaut}. These
contributions are likely to be insignificant within the intrinsic
inaccuracy of the flame speed model

The SGS turbulence stress tensor $\tau_{ik}$ enhances the viscous
dissipation given by $\sigma_{ik}$ in the equation of motion for the
filtered velocity field:
\begin{equation}
  \label{eq:qnse}
  \rho\frac{\DD}{\DD t}v_{i} =
  -\frac{\partial P}{\partial x_{i}} + \rho f_{i}^{(\mathrm{s})} + 
  \frac{\partial}{\partial x_{k}}(\sigma_{ik} + \tau_{ik}).
\end{equation}
Here it is assumed that the specific stirring force
$f_{i}^{(\mathrm{s})}$ injects energy on length scales which are
large compared to the numerical resolution $\Delta$.
The operator $\DD/\DD t$ is the \emph{Lagrangian derivative},
\begin{equation}
  \frac{\DD}{\DD t} = \frac{\partial}{\partial t} +
  \vect{v}\cdot\vect{\nabla}.
\end{equation}

The dynamical equations for kinetic energy in the Germano
decomposition read
\begin{align}
  \label{eq:res_energy}
  \rho\frac{\DD}{\DD t}k_{\mathrm{res}} &= 
     v_{i}\left[-\frac{\partial P}{\partial x_{i}} + \rho f_{i}^{(\mathrm{s})} + 
                \frac{\partial}{\partial x_{k}}\tau_{ik}\right] \\
  \label{eq:sgs_energy}
  \rho\frac{\DD}{\DD t}k_{\mathrm{sgs}} &- \mathfrak{D}_{\mathrm{sgs}} =
  \Sigma_{\mathrm{sgs}} - \rho(\lambda_{\mathrm{sgs}} +
  \epsilon_{\mathrm{sgs}}).
\end{align}
In the first
equation, the rate of viscous dissipation of kinetic energy,
$\vect{\nabla}\cdot(\vect{v}\cdot\vect{\sigma})$, is neglected under
the assumption that the flow is virtually unaffected by microscopic
viscosity on length scales greater than the grid resolution $\Delta$.
This is a valid approximation for $\Delta\gg\eta_{\mathrm{K}}$, where
$\eta_{\mathrm{K}}$ is the Kolmogorov scale of viscous dissipation.
For the numerical simulation discussed later-on, $\Delta\sim
10^{3}\,\mathrm{cm}$, whereas $\eta_{\mathrm{K}}\ll 1\,\mathrm{cm}$
\cite{NieWoos97}.  The symbolic terms in
equation~(\ref{eq:sgs_energy}) account for the diffusion, production
and dissipation of SGS turbulence energy (see \cite{Schmidt04} for the
exact definitions).  Energy transfer from resolved toward subgrid
scales is given by the rate of production $\Sigma_{\mathrm{sgs}}$. The
non-local transport term $\mathfrak{D}_{\mathrm{sgs}}$ accounts for
the redistribution of turbulence energy by subgrid scale velocity and
pressure fluctuations.  Furthermore, there are two contributions to
the rate of dissipation: $\rho\epsilon_{\mathrm{sgs}}$ is caused by
the viscosity of the fluid, while $\rho\lambda_{\mathrm{sgs}}$ is due
to compression effects.  In fact, all of the SGS dynamical terms are
\emph{non-computable} in terms of resolved quantities. This is a
consequence of the non-linear structure of the hydrodynamical
equations, which prohibits complete decomposition. In consequence, one
must find heuristic approximations in terms of computable quantities,
which are commonly known as SGS \emph{closures}.

We apply the customary turbulent-viscosity hypothesis for the rate of
production (section~10.1 in \cite{Pope}), the gradient-diffusion
hypothesis for turbulent transport (section~4.3 in \cite{Sagaut}) and
the dimensional closure for the rate of viscous dissipation
(section~13.6.3 in \cite{Pope}). Since pressure effects are small for
deflagration in degenerate matter, a rather crude closure for
$\lambda_{\mathrm{sgs}}$ is utilised \cite{Dear73}. The final result
is the following dynamical equation (section~3.1.4 in
\cite{Schmidt04}):
\begin{equation}
\begin{split}
  \label{eq:sgs_energy_cl}
  \frac{\DD}{\DD t}k_{\mathrm{sgs}} -& \frac{1}{\rho}
  \vect{\nabla}\cdot\left(\rho C_{\kappa}\Delta_{\mathrm{eff}}
    k_{\mathrm{sgs}}^{1/2}\vect{\nabla}k_{\mathrm{sgs}}\right) = \nonumber\\
  & C_{\nu}\Delta_{\mathrm{eff}}k_{\mathrm{sgs}}^{1/2}|S^{\ast}|^{2} -
  \left(\frac{2}{3}+C_{\lambda}\right)k_{\mathrm{sgs}}d -
  C_{\epsilon}\frac{k_{\mathrm{sgs}}^{3/2}}{\Delta_{\mathrm{eff}}}.
\end{split}
\end{equation}
In particular, an expression analogous to the viscous dissipation term
in the Navier-Stokes equations, $\sigma_{ik}=\rho\nu S_{ik}$ is substituted for the anisotropic part
of $\tau_{ik}$. The rate of SGS turbulence production is then given by
\begin{equation}
  \Sigma_{\mathrm{sgs}} = \tau_{ik}S_{ik} =
  \rho\left(\nu_{\mathrm{sgs}}|S^{\ast}| - \frac{2}{3}k_{\mathrm{sgs}}d\right),
\end{equation}
where $\nu_{\mathrm{sgs}}=
C_{\nu}\Delta_{\mathrm{eff}}k_{\mathrm{sgs}}^{1/2}$ is the SGS
turbulence viscosity.  The \emph{rate-of-strain} tensor $S_{ik}$ is
the symmetrised spatial derivative of the velocity field:
\begin{equation}
  S_{ik} = v_{(i,k)} \equiv 
  \frac{1}{2}\left(\frac{\partial v_{i}}{\partial x_{k}} + 
                   \frac{\partial v_{k}}{\partial x_{i}}\right).
\end{equation}
The trace of this tensor yields the \emph{dilatation} of the velocity
field, $d=S_{ii}$. The scalar $|S^{\ast}|$ is formed by total
contraction of the trace-free part of the rate-of-strain tensor,
\begin{equation}
  \label{eq:strain_norm}
  |S^{\ast}| = \sqrt{2S_{ik}^{\ast}S_{ik}^{\ast}} =
  \sqrt{2\left(S_{ik}S_{ik}-\frac{1}{3}d^{2}\right)}.
\end{equation}
The length scale $\Delta_{\mathrm{eff}}$ is an effective scale of the
finite-volume scheme, namely, the PPM. The ratio
$\beta=\Delta_{\mathrm{eff}}/\Delta$ specifies the smoothing of the
flow on the smallest resolved scales due to numerical dissipation. In
\cite{SchmHille05}, we propose a method of calculating $\beta$ from
numerical realisations of isotropic turbulence.  For moderately
compressible flows, it appears that $\beta\approx 1.6$.

Alternatively, a dynamical equation for the turbulence velocity
$q_{\mathrm{sgs}}=\sqrt{2k_{\mathrm{sgs}}}$ can be formulated:
\begin{equation}
  \label{eq:sgs_velocity}
  \frac{\DD}{\DD t}q_{\mathrm{sgs}} - \frac{1}{\rho}
  \vect{\nabla}\cdot\left(\rho\ell_{\kappa}q_{\mathrm{sgs}}
                   \vect{\nabla}q_{\mathrm{sgs}}\right) -
  \ell_{\kappa}|\vect{\nabla}q_{\mathrm{sgs}}|^{2} =
  \ell_{\nu}|S^{\ast}|^{2} - 
  \left(\frac{1}{3}+\frac{C_{\lambda}}{2}\right)q_{\mathrm{sgs}}d -
  \frac{q_{\mathrm{sgs}}^{2}}{\ell_{\epsilon}}.
\end{equation}
The length scales introduced above are defined as follows:
\begin{equation}
  \ell_{\kappa}=C_{\kappa}\Delta_{\mathrm{eff}}/\sqrt{2}, \qquad
  \ell_{\nu}=C_{\nu}\Delta_{\mathrm{eff}}/\sqrt{2}, \qquad
  \ell_{\epsilon}=2\sqrt{2}\Delta_{\mathrm{eff}}/C_{\epsilon}.
\end{equation}
The equation for $q_{\mathrm{sgs}}$ can be evolved starting with the
initial data $q_{\mathrm{sgs}}(\vect{x},0) = 0$ for a fluid
being initially at rest. Moreover, non-integer powers of
$q_{\mathrm{sgs}}$ do not occur, and the functional dependence on
$q_{\mathrm{sgs}}$ is advantageous for the discretisation of the
diffusion term. However, numerical errors may arise from the
non-conservative form of equation~(\ref{eq:sgs_velocity}). Since we
apply the solution predominantly to estimate the turbulent flame
speed, this caveat is not of much concern.

At this point, one is left with the problem of determining the closure
parameters $C_{\kappa}$, $C_{\nu}$, $C_{\epsilon}$ and
$C_{\lambda}$. For isotropic turbulence, approximate statistical
values from analytic theories or numerical data can be found. We
adopted the constant parameter $C_{\lambda}=-0.2$ \cite{FurTab97}, and
the turbulent diffusion parameter $C_{\kappa}=0.36$ was estimated from
inertial-subrange properties of flow realisations in numerical
simulations of forced isotropic turbulence (section~3.2.4 in
\cite{Schmidt04}).

A more sophisticated approach is the numerical \emph{in situ}
computation of SGS closure parameters from local structural properties
of the flow. The underlying idea is that turbulence in the inertial
subrange becomes asymptotically self-similar towards smaller length
scales. In other words, mostly turbulent velocity fluctuations on the
smallest numerically resolved length scales determine the local energy
transfer towards unresolved scales. This idea initiated the
development of so-called dynamical procedures for the computation of
$C_{\nu}$.  The result is a \emph{localised closure} for the rate of
energy transfer $\Sigma_{\mathrm{sgs}}$ (section~4.3 in
\cite{Sagaut}). For the rate of dissipation,
$\epsilon_{\mathrm{sgs}}$, localised closures have been suggested as
well. Given the computational difficulties and conceptual shortcomings
of these closures, we decided to apply a statistical method for
calculating time-dependent mean dissipation parameters in regions
containing fuel, flames or ash, respectively. In the following
section, we will explain the computational procedures for $C_{\nu}$
and $C_{\epsilon}$ in detail. A generalisation including dynamical
procedures for $C_{\kappa}$ or $C_{\lambda}$ as well would be extremely
involved and is likely to be infeasible in terms of computational
costs. Fortunately, the most important contributions to SGS dynamics
arise from the production and the dissipation terms.

\subsection{The semi-localised model}
\label{sc:sgs_locl}

In order to extract the small scale velocity fluctuations in a
simulation, a \emph{test filter} $\langle\ \rangle_{\mathrm{T}}$ is
applied. This filter smooths the numerically computed flow over a
characteristic length $\Delta_{\mathrm{T}}=\gamma_{\mathrm{T}}\Delta_{\mathrm{eff}}$,
where the factor $\gamma_{\mathrm{T}}>1$.  It is then possible to
compute the turbulence stress associated with the intermediate range
of length scales $\Delta\lessapprox l\lessapprox\Delta_{\mathrm{T}}$:
\begin{equation}
  \tau_{\mathrm{T}}(v_{i},v_{k}) =
  -\langle\rho v_{i}v_{k}\rangle_{\mathrm{T}} +
    \frac{1}{\langle\rho\rangle_{\mathrm{T}}}
    \langle\rho v_{i}\rangle_{\mathrm{T}}\langle\rho v_{k}\rangle_{\mathrm{T}},
\end{equation}
Applying the eddy-viscosity closure to the trace-free part of
$\tau_{\mathrm{T}}(v_{i},v_{k})$, we have
\begin{equation}
  \label{eq:c_prod_test}
  \tau_{\mathrm{T}}^{\ast}(v_{i},v_{k})S_{ik}^{[\mathrm{T}]}
  = {\rho_{\mathrm{T}}C_{\nu}\Delta_{\mathrm{T}}k_{\mathrm{T}}^{1/2}
    |S^{\ast\,[\mathrm{T}]}|^{2}},
\end{equation}
where $\rho_{\mathrm{T}}=\langle\rho\rangle_{\mathrm{T}}$ is the
test-filtered mass density, and the rate of strain at the test filter level
is defined by $S_{ik}^{[\mathrm{T}]}= \partial_{(i}\langle\rho
v_{j)}\rangle_{\mathrm{T}}/\rho_{\mathrm{T}}$.  The
expression $C_{\nu}\Delta_{\mathrm{T}}k_{\mathrm{T}}^{1/2}$ has the
dimension of viscosity. The kinetic energy $k_{\mathrm{T}}$ is defined
analogous to equation~(\ref{eq:sgs_energy_cl}), with the test filter in place
of the implicit filter and the numerically computed velocity
$\vect{v}$ in place of $\overset{\infty}{\vect{v}}$. 

Invoking the similarity hypothesis that $C_{\nu}$ is equal for the
eddy-viscosity closure at the test filter level and for the unfiltered
SGS turbulence stress, the anisotropic part of the rate of production in the localised SGS
model is given by
\begin{equation}
  \label{eq:prod_sgs}
  \frac{1}{\rho}\Sigma_{\mathrm{sgs}} + \frac{2}{3}k_{\mathrm{sgs}}d \circeq 
    \ell_{\nu}q_{\mathrm{sgs}}|S^{\ast}|^{2} =
  \frac{\tau_{\mathrm{T}}^{\ast}(v_{i},v_{k})S_{ik}^{[\mathrm{T}]}}
       {\gamma_{\mathrm{T}}\rho_{\mathrm{T}}}\,
    \frac{|S^{\ast}|^{2}}{|S^{\ast\,[\mathrm{T}]}|^{2}}
    \sqrt{\frac{k_{\mathrm{sgs}}}{k_{\mathrm{T}}}}.
\end{equation}
The above closure is basically the result of adapting the
Germano-Lilly dynamical procedure for the localised Smagorinsky model
to the SGS turbulence energy model \cite{GerPio91}. As an important
difference, however, the eddy viscosity closure is applied to
$\tau_{\mathrm{T}}(v_{i},v_{k})$ rather than the total turbulence
stress at the test filter level,
$\tau_{\mathrm{T}}(\overset{\infty}{v_{i}},\overset{\infty}{v_{k}})$.
The stress tensors are related by the Germano identity
\cite{Germano92}:
\begin{equation}
  \label{eq:germano}
  \tau_{\mathrm{T}}(\overset{\infty}{v_{i}},\overset{\infty}{v_{k}}) =
  \langle\tau(\overset{\infty}{v_{i}},\overset{\infty}{v_{k}})\rangle_{\mathrm{T}} +
  \tau_{\mathrm{T}}(v_{i},v_{k})
\end{equation}
This modification of the dynamical procedure was proposed by Kim
\textit{et al.} \cite{KimMen99}. It is supported by results from the
evaluation of velocity measurements in round jets \cite{LiuMen94} and
was explicitly verified on data from simulations of compressible
turbulence driven by stochastic stirring (section~3.2.2 in \cite{Schmidt04}).

A complication arises from $C_{\nu}$ becoming negative in some regions
of a turbulent flow. This is commonly interpreted as
\emph{backscattering}, i.e.\ kinetic energy is locally transfered
across the cutoff from smaller, unresolved vortices towards vortices
of size larger than $\Delta$ (section~4.4 in \cite{Sagaut}). Including
the contributions from backscattering in numerical simulations
introduces several difficulties \cite{Pio93}. Firstly, numerical
instabilities might be induced, because backscattering amounts to
negative diffusion.  Secondly, the SGS turbulence stresses must be
coupled to the resolved flow in order to consistently account for the
conversion of SGS turbulence energy into resolved kinetic energy.
This is exactly what one would do in conventional large eddy
simulations. However, in combination with a dissipative finite-volume
scheme such as the PPM, including the SGS stress terms in the momentum
equation does not do much good.  Inevitably, the kinetic energy
produced by positive SGS stress (corresponding to negative SGS
viscosity) would be injected into modes corresponding to wave numbers
near the cutoff. These wave numbers, however, are severely affected by
numerical dissipation \cite{SchmHille05}, and the fluid motion
produced by the inverse energy transfer would be rapidly damped
out. Consequently, backscattering would effectively result in enhanced
dissipation, i.e.\ conversion of subgrid scale turbulence energy into
internal rather than resolved kinetic energy.  For this reason, the
outcome of suppressing backscattering will be investigated in
section~\ref{sc:turb_burn_sim}.

Finding a dynamical procedure for the parameter of SGS dissipation,
$C_{\epsilon}$, is yet more demanding. The difficulty of determining
$C_{\epsilon}$ stems from the fact the rate of dissipation is mostly
determined by the fluid dynamics on scales much smaller than the
numerical resolution. Therefore, a localised similarity hypothesis is
bound to fail. However, one can invoke a statistical
argument. Considering a certain region of the flow, the average rate
of dissipation in that region should be roughly balanced by the mean
transfer from larger toward smaller scales, if the flow is nearly in
statistical equilibrium. Even in developing flows, a time-dependent
statistical value of $C_{\epsilon}$ can be calculated by means of
energy conservation.  The method is loosely based on the variational
approach of \cite{GhoLund95}, where the parameter of dissipation
$C_{\epsilon}$ is determined by subtracting the test-filtered SGS
turbulence energy equation~(\ref{eq:sgs_energy}) from the
corresponding equation for the unresolved kinetic energy at the level
of the test filter.  Rather than computing $C_{\epsilon}$ locally, we
will determine statistical values evolving in time from the spatially
averaged energy equations. Upon averaging
equation~(\ref{eq:sgs_energy}), one obtains
\begin{equation}
  \label{eq:energy_sgs_ave}
  \left\langle\rho\frac{\DD}{\DD t}k_{\mathrm{sgs}}\right\rangle = 
  \langle\tau_{ik}S_{ik}\rangle - 
  \left\langle\rho(\lambda_{\mathrm{sgs}}+\epsilon_{\mathrm{sgs}})\right\rangle.
\end{equation}
The diffusion term cancels out, because integrating the divergence of
the diffusive flux over a domain with periodic BCs yields zero. 
Furthermore,
\begin{equation}
  \left\langle\rho\frac{\DD}{\DD t}k_{\mathrm{sgs}}\right\rangle =
  \left\langle\frac{\partial}{\partial t}\rho k_{\mathrm{sgs}}\right\rangle +
  \underbrace{\left\langle\frac{\partial}{\partial x_{i}}
              \rho v_{i}k_{\mathrm{sgs}}\right\rangle}_{=0} =
  \frac{\dd}{\dd t}\langle K_{\mathrm{sgs}}\rangle,
\end{equation}
i.e.\ there is vanishing net advection over the whole domain of
the flow. The turbulence energy at the characteristic scale of the
test filter is defined by
\begin{equation}
  \label{eq:energy_sgs_test_ave}
  -\frac{1}{2}\tau_{\mathrm{T}}(\overset{\infty}{v_{i}},\overset{\infty}{v_{i}}) =
  -\frac{1}{2}\langle\tau_{ii}\rangle_{\mathrm{T}} + \frac{1}{2}\tau_{\mathrm{T}}(v_{i},v_{i}) =
  \langle\rho k_{\mathrm{sgs}}\rangle_{\mathrm{T}} + \rho_{\mathrm{T}} k_{\mathrm{T}},
\end{equation}
and the corresponding averaged dynamical equation is
\begin{equation}
  \frac{\partial}{\partial t}\langle
    \rho K_{\mathrm{sgs}}+\rho_{\mathrm{T}}K_{\mathrm{T}}\rangle = 
  \left\langle\tau_{\mathrm{T}}(\overset{\infty}{v_{i}},\overset{\infty}{v_{k}})
         S_{ik}^{[\mathrm{T}]}\right\rangle - 
  \left\langle\rho(\lambda_{\mathrm{sgs}}+\epsilon_{\mathrm{sgs}})+
         \rho_{\mathrm{T}}(\lambda_{\mathrm{T}}+\epsilon_{\,\mathrm{T}})\right\rangle.
\end{equation}
Equations~(\ref{eq:energy_sgs_ave})
and~(\ref{eq:energy_sgs_test_ave}) in combination with the Germano
identity~(\ref{eq:germano}) imply the following conservation law for the mean turbulence
energy $\langle K_{\mathrm{T}}\rangle$ associated with the smallest resolved scales:
\begin{equation}
  \label{eq:energy_test_ave}
  \frac{\dd}{\dd t}\langle K_{\mathrm{T}}\rangle = 
  \left\langle\tau_{\mathrm{T}}(v_{i},v_{k})S_{ik}^{[\mathrm{T}]} +
         \langle\tau_{ik}\rangle_{\mathrm{T}}S_{ik}^{[\mathrm{T}]} -
         \tau_{ik}S_{ik}\right\rangle - 
  \left\langle\rho_{\mathrm{T}}(\lambda_{\mathrm{T}}+\epsilon_{\,\mathrm{T}})\right\rangle.
\end{equation}

Substituting the turbulent-viscosity closures for the various
production terms on the right-hand side, the above equation becomes
\begin{equation}
  \begin{split}	
     \frac{\dd}{\dd t}\langle K_{\mathrm{T}}\rangle \simeq
     & \underbrace{\left\langle\rho_{\mathrm{T}} C_{\nu}\Delta_{\mathrm{T}}\sqrt{k_{\mathrm{T}}}\,
                  |S^{\ast\,[\mathrm{T}]}|^{2}\right\rangle -
                \frac{2}{3}\left\langle K_{\mathrm{T}}
                  d^{[\mathrm{T}]}\right\rangle}_{\mathrm{(I)}} - 
      \langle\rho_{\mathrm{T}}\lambda_{\mathrm{T}}\rangle +
      \langle\rho_{\mathrm{T}}\epsilon_{\,\mathrm{T}}\rangle \nonumber\\
     & + \underbrace{\left\langle\langle\rho\nu_{\mathrm{sgs}}S_{ik}^{\ast}\rangle_{\mathrm{T}}
                  S_{ik}^{\ast\,[\mathrm{T}]} -
                \rho\nu_{\mathrm{sgs}}|S^{\ast}|^{2}\right\rangle}_{\mathrm{(II)}} -
      \frac{2}{3}\underbrace{\left\langle\langle K_{\mathrm{sgs}}\rangle_{\mathrm{T}}d^{[\mathrm{T}]}-
                  K_{\mathrm{sgs}}d\right\rangle}_{\mathrm{(III)}}.
  \end{split}                 
\end{equation}
Analogous to the rate of strain at the test filter level, the
divergence $d^{[\mathrm{T}]}$ is given by $d^{[\mathrm{T}]}=
\partial_{i}\langle\rho v_{i}\rangle_{\mathrm{T}}/\rho_{\mathrm{T}}$.
The most significant production term is (I) which measures the energy
transfer across the test filter scale $\Delta_{\mathrm{T}}$. The two
contributions in term (II), on the other hand, are both related to the
energy transfer across $\Delta_{\mathrm{eff}}$, where the first
expression is calculated from the test-filtered and the second
expression from the numerically resolved rate of strain,
respectively. It appears reasonable to assume that the difference of
these two expressions is marginal relative to (I) in the case of
scaling ratios $\Delta_{\mathrm{T}}/\Delta_{\mathrm{eff}}$ of the
order unity.  Furthermore, the evaluation of (II) is particularly
costly due to several tensor components which have to be
test-filtered. Thus, we neglect term (II). For similar reasons and
because of the smallness of compressibility effects, we drop (III) as
well.  In conclusion, the rate of dissipation
$\epsilon_{\,\mathrm{T}}$ is approximately given by
\begin{equation}
  \label{eq:energy_test_ave_approx}
  \langle\rho_{\mathrm{T}}\epsilon_{\,\mathrm{T}}\rangle \simeq
  -\frac{\dd}{\dd t}\langle K_{\mathrm{T}}\rangle +
  \left\langle\rho_{\mathrm{T}} C_{\nu}\Delta_{\mathrm{T}}\sqrt{k_{\mathrm{T}}}\,
    |S^{\ast\,[\mathrm{T}]}|^{2}\right\rangle -
  \frac{2}{3}\left\langle\rho_{\mathrm{T}}(k_{\mathrm{T}}d^{[\mathrm{T}]}+
    \lambda_{\mathrm{T}})\right\rangle.
\end{equation}

The crucial step is to conjecture that the relation between the
spatially averaged dissipation rate and turbulence energy is similar
at the cutoff and the test filter level. This implies
\begin{equation}
  \label{eq:diss_test1}
  \langle\rho_{\mathrm{T}}\epsilon_{\,\mathrm{T}}\rangle \circeq
  \frac{C_{\mathrm{\epsilon}}}{\Delta_{\mathrm{T}}}\left\langle\rho_{\mathrm{T}}
    \left(\frac{\langle\rho k_{\mathrm{sgs}}\rangle_{\mathrm{T}}}{\rho_{\mathrm{T}}} +
          k_{\mathrm{T}}\right)^{3/2} -
    \gamma_{\mathrm{T}}\rho k_{\mathrm{sgs}}^{3/2}\right\rangle.
\end{equation}
Note that the expression in parentheses is the total turbulence energy
at the test filter level. For the pressure-dilatation term
$\lambda_{\mathrm{T}}$, we set
\begin{equation}
  \label{eq:pd_test}
  \lambda_{\mathrm{T}}\circeq 
  C_{\mathrm{\lambda}}k_{\mathrm{T}}d^{[\mathrm{T}]},
\end{equation}
which is analogous to the closure at the subgrid-scale level.  The
rate of SGS dissipation is therefore given by
\begin{equation}
\begin{split}
  \label{eq:c_diss_ave}
  \epsilon_{\mathrm{sgs}} =
  & -\gamma_{\mathrm{T}}\left\langle\rho_{\mathrm{T}}
  \left(\frac{\langle\rho k_{\mathrm{sgs}}\rangle_{\mathrm{T}}}{\rho_{\mathrm{T}}} +
        k_{\mathrm{T}}\right)^{3/2} -
  \gamma_{\mathrm{T}}\rho k_{\mathrm{sgs}}^{3/2}\right\rangle^{-1}\\
  & \times\left[\frac{\dd}{\dd t}\langle K_{\mathrm{T}}\rangle - 
  \left\langle C_{\nu}\rho_{\mathrm{T}}\Delta_{\mathrm{T}}\sqrt{k_{\mathrm{T}}}\,
          |S^{\ast\,[\mathrm{T}]}|^{2}\right\rangle +
  \left(\frac{1}{3}+C_{\lambda}\right)\left\langle K_{\mathrm{T}}
                                                  d^{[\mathrm{T}]}\right\rangle\right]
  k_{\mathrm{sgs}}^{3/2}.
\end{split}
\end{equation}
As opposed to the statistical values for the SGS parameters for steady
isotropic turbulence, the above equation yields a spatially constant
parameter evolving in time. This method of calculating $C_{\epsilon}$
in combination with the dynamical procedure for $C_{\nu}$ makes up the
\emph{semi-localised} SGSTE model.  For the numerical implementation,
two further modifications were added.

On account of the anisotropy in the vicinity of a flame front, it
seems advisable to average over the principal topological subdomains,
namely, the interior, the exterior and the interface. The latter is
identified by marking all grid cells within a certain maximal distance
to those cells in which the level set function $G$ swaps its sign.
With this procedure, the functions $C_{\epsilon}^{(\mathrm{a})}(t)$,
$C_{\epsilon}^{(\mathrm{b})}(t)$ and $C_{\epsilon}^{(\mathrm{f})}(t)$
are obtained for the mean dissipation parameters in ash, the burning
zone and fuel, respectively.  In the early stage, burning regions
might encompass only a small volume fraction with relatively high
surface to volume ratio. Hence, the corresponding spatial averages in
equation~(\ref{eq:c_diss_ave}) will not be sufficiently well behaved
at the beginning.  Although, the dynamics is dominated by the fuel
domain at this point, both the enumerator and dominator in
equation~(\ref{eq:c_diss_ave}) are smoothed in time via convolution
with an exponential damping function in order to remove strong
oscillations in the ash and flame regions. The characteristic time
scale of smoothing is prescribed by the parameter $T_{\epsilon}$.  An
appropriate choice for the time scale $T_{\epsilon}$ has to be found
\emph{a posteriori}. Setting $T_{\epsilon}\approx 0.1T$ appears to be
a good choice in order to get well behaved functions
$C_{\epsilon}^{(\mathrm{a})}(t)$, $C_{\epsilon}^{(\mathrm{b})}(t)$ and
$C_{\epsilon}^{(\mathrm{f})}(t)$, without overly damping dynamical
variations (section~4.3.4 in \cite{Schmidt04}).

\section{Numerical Simulations}

The simplest case study one can think of is the evolution of flame
fronts in spatially homogeneous turbulent flows. To that end, we
implemented a stochastic driving mechanism for the production of
turbulence in a cubic domain subject to periodic boundary conditions
\cite{Schmidt04,EswaPope88}.  The notion of a stochastic force field
is outlined below. The computational domain is divided into a lattice
of subcubes of size $L=X/\alpha$, where $L$ is the characteristic
wavelength of the stochastic driving force and $X$ is the domain
size. In the centre of each subcube, thermonuclear burning is ignited
in small spherical regions located at time $t=0$, when the stirring
force begins to act on the fluid. We chose $\alpha=2$, giving eight
subcubes in the computational domain. This pattern is infinitely
repeated in space by virtue of the periodic boundary conditions.
Gravity is negligible at the scales under consideration (section~2.3.3
in \cite{Schmidt04}). Consequently, there are no buoyancy effects
and turbulence is only produced by stirring.

The crucial parameter for the evolution of the burning process is the
ratio $\xi=s_{\mathrm{lam}}/V$, i.e.\ the ratio of the laminar
burning speed to the characteristic velocity of the turbulent flow.
Assuming developed turbulence, one can apply the Kolmogorov scaling
law and estimate the magnitude of turbulent velocity fluctuations at a
separation of the order to the Gibson length:
\begin{equation}
  v'(l_{\mathrm{G}}) \sim V\left(\frac{l_{\mathrm{G}}}{L}\right)^{1/3}.
\end{equation}
The integral length $L$ and characteristic velocity $V$ specify
the largest turbulent vortices in the flow. Setting
$v'(l_{\mathrm{G}})=s_{\mathrm{lam}}$, the scaling law for the
Gibson length becomes
\begin{equation}
  \label{eq:gibson_intgr}
  l_{\mathrm{G}}\sim L\left(\frac{s_{\mathrm{lam}}}{V}\right)^{3}=L\xi^{3}
\end{equation}
Obviously, $l_{\mathrm{G}}$ is very sensitive to value of $\xi$.
Possible choices of $V$ are restricted by the speed of sound
$c_{\mathrm{s}}$.  Both $c_{\mathrm{s}}$ and $s_{\mathrm{lam}}$ are
mostly determined by the mass density, and so is the ratio
$l_{\mathrm{G}}/L$. In the following, we will consider two distinct
cases.  For $\xi>0.1$ and sufficiently high resolution, the Gibson
scale is just within the range of numerically resolved length
scales. In this case, no SGS model is required for the flame dynamics
and the propagation speed is more or less given by the laminar flame
speed. If $\xi\ll0.1$, on the other hand, it is impossible to resolve
the flame completely. Then the burning process will enter a turbulent
regime, in which the flame propagation speed is asymptotically given
the SGS turbulence velocity $q_{\mathrm{sgs}}$.  Prior to the
discussion of the numerical simulations, we give a brief description of
the stirring mechanism for the production of isotropic turbulent flow.

\subsection{Stochastic forcing}

The specific driving force $\vect{f}(\vect{x},t)$ is composed in spectral
space, using a three-dimensional generalisation of the scalar
\emph{Ornstein-Uhlenbeck process}, as proposed in \cite{EswaPope88}.
The evolution of the Fourier transform $\hat{\vect{f}}(\vect{k},t)$ is
given by the following Langevin-type stochastic differential equation:
\begin{equation}
  \label{eq:stirr_mode_evol}
  \dd\hat{\vect{f}}(\vect{k},t) = 
  -\hat{\vect{f}}(\vect{k},t)\frac{\dd t}{T} +
  F_{0}\sum_{jlm}\left(\frac{2\sigma^{2}(\vect{k})}{T}\right)^{1/2}
    \delta(\vect{k}-\vect{k}_{jlm})
    \tens{P}_{\zeta}(\vect{k})\cdot\dd\boldsymbol{\mathcal{W}}_{t},
\end{equation}
The second term on the right hand side accounts for a random diffusion
process, which is constructed from a three-component \emph{Wiener
process} $\boldsymbol{\mathcal{W}}_{t}$. The distribution of each component
is normal with zero mean and variance $\dd t$. The wave vectors
$\vect{k}_{jlm}$ are dual to the position vectors of the cells in the
numerical discretisation of the fundamental domain.  The symmetric
tensor $\tens{P}_{\zeta}(\vect{k})$ is defined by the linear
combination of the projection operators perpendicular and parallel to
the wave vector. The components of
$\tens{P}_{\zeta}(\vect{k})$ can be expressed as
\begin{equation}
  ({\sf P}_{ij})_{\zeta}(\vect{k}) = 
  \zeta {\sf P}_{ij}^{\perp}(\vect{k}) + 
  (1-\zeta){\sf P}_{ij}^{\parallel}(\vect{k}) =
  \zeta\delta_{ij} + (1-2\zeta)\frac{k_{i}k_{j}}{k^{2}},
\end{equation}
where the spectral weight $\zeta$ determines whether the resulting
force field in physical space is purely solenoidal, dilatational or a
combination of both. The variance $\sigma^{2}(\vect{k})$ specifies the
spectrum of the force field. We use a quadratic function, which
confines the modes of the force to a narrow interval of wavenumbers,
$k\in[0,2k_{0}]$. The wave number $k_{0}$ determines the
\emph{integral length scale} of the flow, $L=2\pi/k_{0}$.

The root mean square of the specific driving force is determined by
the characteristic magnitude $F_{0}$ and the weight $\zeta$:
\begin{equation}
  \label{eq:stirr_rms}
  f_{\mathrm{rms}} = 
  \sum_{jlm}\langle\hat{\vect{f}}_{jlm}(t)\cdot\hat{\vect{f}}_{jlm}(t)\rangle
  \simeq (1 - 2\zeta + 3\zeta^{2})F_{0}^{2}.
\end{equation}
Since $F_{0}$ has the physical dimension of acceleration, it can be
expressed as the characteristic velocity $V$ of the flow
divided by the integral time scale $T$, which is given by the
auto-correlation time $T$ of the driving force
(\ref{eq:stirr_mode_evol}). Setting $T=L/V$, we have $F_{0}=V/T=L
V^{2}$, and, starting with a homogeneous fluid at rest, the flow is
developing towards a fully turbulent steady state within about two
integral time scales.

\subsection{Quasi-laminar burning}
\label{sc:lam_burn_sim}

To begin with, we shall consider the case $\xi\sim 1$. Then the
laminar flame propagation is fast enough to burn the smallest
numerically resolved eddies in less than
 a turn-over time. Consequently,
subgrid scale turbulence does not affect the flame dynamics. An
estimate of the characteristic velocity $V$ for given numerical
resolution and laminar burning speed is readily obtained from
relation~(\ref{eq:gibson_intgr}).  The effective range of length
scales which can be resolved is roughly given by
$L/\Delta_{\mathrm{eff}}=N/\alpha\beta$, where $N=\Delta/X$ is the
number of numerical cells in one dimension.  For
$l_{\mathrm{G}}\approx\Delta_{\mathrm{eff}}$, we therfore must have
\begin{equation}
 \label{eq:max_intgr_vel}
 V\approx\left(\frac{N}{\alpha\beta}\right)^{1/3}s_{\mathrm{lam}}.
\end{equation}
For the simulation, which will be discussed in the following, we used
$432^{3}$ grid cells. Setting $N=432$,
relation~(\ref{eq:max_intgr_vel}) implies that at most $V\approx
4s_{\mathrm{lam}}$ is admissible.  Given a moderate mass density, this
would entail an extremely low Mach number. However, computing an
almost incompressible flow with the PPM would be infeasible. On the
other hand, for an initial density $\rho_{0}\approx
2.90\cdot10^{9}\,\mathrm{g\,cm^{-3}}$, one obtains
$s_{\mathrm{lam}}\approx 1.05\cdot 10^{7}\,\mathrm{cm\,s^{-1}}\,$
through interpolation of numerical data taken from Timmes and Woosley
\cite{TimWoos92}. The speed of sound for this density is $c_{0}\approx
9.70\cdot 10^{8}\,\mathrm{cm\,s^{-1}}$.  Choosing
$V=4s_{\mathrm{lam}}\approx 4.20\cdot 10^{7}\,\mathrm{cm\,s^{-1}}\,$,
the characteristic Mach number is $V/c_{0}\approx 0.043$. This is
quite small, but still computationally manageable with a fully
compressible hydro code.  The resulting Gibson length,
$l_{\mathrm{G}}\approx 3.3\Delta$, allows for some margin between
$l_{\mathrm{G}}$ and $\Delta_{\mathrm{eff}}\approx 1.6\Delta$.

Actually, Landau-Darrieus instabilities would induce a small-scale cellular flame
structure \cite{RoepHille04a}. Due to the significant numerical
dissipation at length scales $l\lesssim 10\Delta$ \cite{SchmHille05}, the
cellular structure will inevitably get smeared out if
$l_{\mathrm{G}}\sim\Delta$.  For this reason, an effective cellular
propagation speed $s_{\mathrm{cell}}$ slightly larger than
$s_{\mathrm{lam}}$ would be the correct intrinsic propagation speed in
place of the laminar burning speed \cite{NieWoos97}.  However, this
effect is ignored, as the change of the Gibson scale due to the
difference between $s_{\mathrm{lam}}$ and $s_{\mathrm{cell}}$ is
only about a factor of two. Consequently, using $s_{\mathrm{cell}}$
as intrinsic propagation speed would not change the flame dynamics
dramatically.

Regarding the numerical distortion introduced by the passive
implementation, one should be on the safe side for mass densities
larger than $10^{8}\,\mathrm{g\,cm^{-3}}$. Apart from that, the
randomisation caused by turbulence tends to diffuse any numerical
artifacts. In this respect, propagating symmetric flame fronts, say,
nearly planar or spherical flames, is a more demanding
task. Furthermore, flow maps prepared form the simulation data clearly
show a tight correlation between the shape of the front and the flow
structure (see figure~\ref{fg:dns_strain_evol}).  If there were
significant spurious propagation or deformation, the evolution of the
front should become increasingly uncorrelated to the flow. In
conclusion, the simulations which will be discussed subsequently are
likely to give a sound description of the flame dynamics, albeit the
shortcomings of the level set method in the passive implementation.

\begin{figure}
  \begin{center}
    \vspace{110mm}
    \texttt{dns\_energy\_evol.png}
    \bigskip
    \caption{Three-dimensional simulation of thermonuclear
      deflagration in a cube with the flame propagation speed being
      equal to the laminar burning speed. The initial density of the
      C+O fuel is $\rho_{0}\approx
      2.90\cdot10^{9}\,\mathrm{g\,cm^{-3}}$, and
      $V=4s_{\mathrm{lam}}$ ($\xi=0.25$). The characteristic Mach
      number of the fully developed turbulent flow is $V/c_{0}\approx
      0.043$, where $c_{0}$ is the initial sound speed.  Shown are
      two-dimensional contour sections of the normalised specific
      energy $\tilde{e}=e/c_{0}^{2}$ at different stages of the
      burning process.  }
    \label{fg:dns_energy_evol}
  \end{center}
\end{figure}

\begin{figure}
  \begin{center}
    \vspace{110mm}
    \texttt{dns\_densty\_evol.png}
    \bigskip
    \caption{2D contour sections of the relative density fluctuations
      $(\rho-\rho_{0})/\rho_{0}$ corresponding to the panels shown in
      figure~\ref{fg:dns_energy_evol}. }
    \label{fg:dns_densty_evol}
  \end{center}
\end{figure}

\begin{figure}
  \begin{center}
    \vspace{110mm}
    \texttt{dns\_strain\_evol.png}
    \bigskip
    \caption{2D contour sections of the logarithmic dimensionless rate
      of strain, $\log_{10}(T|S^{\ast}|)$, corresponding to the panels
      shown in figures~\ref{fg:dns_energy_evol} and~\ref{fg:dns_densty_evol}. }
    \label{fg:dns_strain_evol}
  \end{center}
\end{figure}

\begin{figure}
  \begin{center}
    \vspace{110mm}
    \texttt{dns\_stat3d.png}
    \bigskip
    \caption{ Evolution of dimensionless statistical moments for the
      simulation illustrated in the series of
      figures~\ref{fg:dns_energy_evol}, \ref{fg:dns_densty_evol}
      and~\ref{fg:dns_strain_evol}.  The top panels show plots of the RMS
      force, momentum and Mach number as well as the average nuclear energy
      generation rate in combination with the chemical composition.
      Furthermore, a measure for the mean flame speed and the averaged
      structural invariants of the flow are plotted in the panels at
      the bottom. }
    \label{fg:dns_stat3d_mean}
  \end{center}
\end{figure}

The progression of the deflagration in the course of the simulation is
illustrated by the sequences of contour plots for the specific internal energy,
the mass density and the rate-of-strain scalar in
figures~\ref{fg:dns_energy_evol}, \ref{fg:dns_densty_evol}
and~\ref{fg:dns_strain_evol}, respectively. The contours of the zero
level set are visible as thin white lines which separate dark regions
containing unburned matter of low energy density from the brightly
coloured regions containing processed material of high energy
density. In the course of the first integral time $T$, the regions of
burned material are expanding gradually. At the same time, they are
stretched an folded by the solenoidal large-scale flow. In the
following second integral time, vortices are generated on
small scales. Subsequent to the peak of $\langle B\rangle$ at
$\tilde{t}=t/T\approx 1.35$, ash encloses fuel rather than the other way
around, and the density of the enclosed fuel is noticeably larger than
the average density. Around $\tilde{t}\approx 1.5$ most of the
fuel has already been consumed by the burning process, and the last
fuel patches are disappearing quickly. Thus, the peak of burning is
reached before turbulence is fully developed and the front propagation
is affected only little by small-scale velocity fluctuations in this
simulation. Consequently, we refer to this mode of burning as
\emph{quasi-laminar}.

In panel (b) of figure~\ref{fg:dns_stat3d_mean}, the mean burning rate
$\langle B\rangle$ is plotted on a dimensionless scale. One can see
that the rate of burning increases exponentially in the interval
$0.3\lesssim\tilde{t}\lesssim 1.2$. The norm $|S^{\ast}|$ of the
trace-free rate of strain defined in equation~(\ref{eq:strain_norm})
is a so-called structural invariant of the flow. Contour plots of
$|S^{\ast}|$ are shown in figure~\ref{fg:dns_strain_evol}. Regions
which are subject to intense strain correspond to steep velocity
gradients and appear bright in the contour plots.  These regions tend
to form vortices and clearly influence the morphology of the flame
fronts.  As one can see in figure~\ref{fg:dns_strain_evol}, the white
lines indicating the zero level set tend to be aligned with structures
associated with large strain. The statistics of $|S^{\ast}|$ and two
further structural invariants, namely, the vorticity
$\omega=|\vect{\nabla}\times\vect{v}|$ and the divergence
$d=\vect{\nabla}\cdot\vect{v}$, is plotted in panel (d) of
figure~\ref{fg:dns_stat3d_mean}.  Because of the small characteristic
Mach number of the flow, we have $d\ll\omega\simeq|S^{\ast}|$, where
the equality of vorticity and rate-of-strain scalar holds
asymptotically in the limit of incompressible flow. The graphs show
that the root mean square (RMS) of $|S^{\ast}|$, grows exponentially
from the first few tenths of an integral time scale up to $t/T\approx
2$, where the stagnation of the growth marks fully developed
turbulence. A comparison between panels (b) and (d) suggests that the
growth of the burning rate prior to the maximum correlates with the
exponentially increasing $\langle|S^{\ast}|^{2}\rangle^{1/2}$. This
underlines the above statement about the influence of strain onto the
flame evolution. The corresponding evolution of the RMS momentum and
Mach number is plotted in panel (a).

The rate of change of the mean mass fraction of fuel,
$\langle\dot{X}(\mathrm{C+O})\rangle$, is also a measure of the
burning speed. For an energy release $\epsilon_{\mathrm{nuc}}$ per
unit mass, the total nuclear energy generated in the whole cubic
domain per unit time can be expressed as
\begin{equation}
  \label{eq:burn_tot1}
  (\alpha L)^{3}\langle B\rangle = 
  -\epsilon_{\mathrm{nuc}}M_{\mathrm{cub}}\langle\dot{X}(\mathrm{C+O})\rangle,
\end{equation}
where $M_{\mathrm{cub}}=\rho_{0}(\alpha L)^{3}$ is the total mass
contained in the computational domain.  On the other hand, the rate of
energy production is related to the total surface area of the
flames, $A_{\mathrm{F}}$, and the laminar propagation speed, 
provided that compression effects are neglected:
\begin{equation}
  \label{eq:burn_tot2}
  (\alpha L)^{3}\langle B\rangle = 
  \rho_{0}\epsilon_{\mathrm{nuc}}A_{\mathrm{F}}s_{\mathrm{lam}}.
\end{equation}
Combining equations~(\ref{eq:burn_tot1}) and~(\ref{eq:burn_tot2}) with
$M_{\mathrm{cub}}=8\rho_{0}L^{3}$, the approximate total surface area
is given by
\begin{equation}
  A_{\mathrm{F}} \simeq
  \frac{(2L)^{3}}{s_{\mathrm{lam}}}\langle\dot{X}(\mathrm{C+O})\rangle.
\end{equation}
The graph of the normalised surface area,
\begin{equation}
  \label{eq:fl_area}
  \tilde{A}_{\mathrm{F}} =
  \frac{A_{\mathrm{F}}}{8\pi^{2}L^{2}} =
  -\frac{1}{\pi^{2}}\frac{V}{s_{\mathrm{lam}}}
    \langle T\dot{X}(\mathrm{C+O})\rangle,
\end{equation} 
is shown in the panel (c) of
figure~\ref{fg:dns_stat3d_mean}. The exponential growth of the burning
rate is manifest in this plot as well. At the peak,
$\tilde{A}_{\mathrm{F}}\sim 1$, which verifies that the flames
experience only little wrinkling due to small vortices.  This result
agrees with the impression of rather smooth flames in
figure~\ref{fg:dns_energy_evol}. Also plotted is the graph of
$-\langle T\dot{X}(\mathrm{C+O})\rangle/\langle
X(\mathrm{C+O})\rangle$, which is a measure of the ratio of the flame
surface area to the amount of still unburned material.\footnote{
Strictly, the
volume of fuel left at a certain time would be given by $(\alpha
L)^{3}\langle\rho X(\mathrm{C+O})\rangle/\rho_{0}$. However, the
mass-weighted fraction of C+O was not calculated in the simulation.}
The mean of this ratio follows a nearly exponential law even beyond
the peak of $\langle B\rangle$ and, thus, can
be considered as an invariant measure for the burning intensity
even when the flames are already diminishing.

\subsection{Burning dominated by turbulence}
\label{sc:turb_burn_sim}

In the case $\xi\ll 1$, i.e.\ the laminar burning speed is small
compared to the chacteristic velocity of the flow, the range of
length scales between the Gibson scale and the integral length scale
becomes very large. Consequently, it is impossible to resolve the
flame dynamics completely. For example, setting $\rho_{0}\approx
2.90\cdot 10^{8}\,\mathrm{g\,cm^{-3}}$, which is by an order of a
magnitude smaller than the density chosen in
section~\ref{sc:lam_burn_sim}, yields $s_{\mathrm{lam}}\approx
9.78\cdot 10^{5}\,\mathrm{cm\,s^{-1}}$. Choosing a characteristic
velocity $V=100s_{\mathrm{lam}}$, we have the Mach number
$V/c_{0}\approx 0.15$, and the Gibson scale becomes
$l_{\mathrm{G}}\sim 10^{-6}L$.  Obviously, $l_{\mathrm{G}}\ll\Delta$
for any feasible numerical resolution.  A subgrid scale model is
therefore mandatory. In this section, several simulations of
thermonuclear deflagration in the cube with the turbulent flame speed
given by equations~(\ref{eq:sgs_flame_speed_max})
and~(\ref{eq:sgs_flame_speed_pocheau}), respectively, are
discussed. The SGS turbulence velocity $q_{\mathrm{sgs}}$ is computed
via equation~(\ref{eq:sgs_velocity}). The initial mass density and the
characteristic velocity are as specified above. Otherwise, the same
parameters as in section~\ref{sc:lam_burn_sim} are used. However, the
resolution is reduced by a factor two. So there are $216^{3}$ grid
cells.

A couple of simulations were performed with the semi-localised model
and the maximum relation~(\ref{eq:sgs_flame_speed_max}) with
$C_{\mathrm{t}}=1$. In one case, we coupled the SGS stresses to the
resolved flow and included backscattering, whereas no coupling was
applied and backscattering was suppressed by setting the SGS viscosity
parameter equal to $C_{\mathrm{\nu}}^{+}=\max(0,C_{\nu})$ in the other
case. The validity of neglecting the SGS stress terms in the momentum
equation~(\ref{eq:qnse}) has been investigated in several
hydrodynamical simulations with PPM \cite{SyPort00,SchmHille05}. For
combustion problems, the SGS model then runs in a passive mode and
provides the turbulent flame speed only.

If negative values of $C_{\nu}$ are admissible, however, the SGS
stress terms must be included, because otherwise backscattering would
convert SGS turbulence energy into heat rather than kinetic energy on
resolved scales. Hence, backscattering necessitates an active SGS
model. For a fully consistent treatment, the terms
$\vect{v}\cdot(\vect{\nabla}\cdot\vect{\tau})$ and
$\rho\epsilon{_\mathrm{sgs}}$ have to be added on the right hand side
of the conservation law for the total energy
$e_{\mathrm{tot}}=\frac{1}{2}|\vect{v}|^{2}+e_{\mathrm{int}}$,
which account for the transfer of kinetic energy between resolved and
subgrid scales and the production of internal energy due to the
viscous dissipation of SGS turbulence energy, respectively.

On the other hand, if backscattering is suppressed, a considerably
simplified scheme is applied, where the dissipation of kinetic energy
is solely of numerical origin, and $q_{\mathrm{sgs}}$ is treated as a
passive scalar. In order to account for the exchange of energy
between the resolved total energy,
$e_{\mathrm{tot}}=\frac{1}{2}v^{2}+e_{\mathrm{int}}$, and the SGS
turbulence energy in a rudimentary fashion, the Lagrangian rate of
change of $k_{\mathrm{sgs}}$ is subtracted from the the
conservation law for the total energy.  Locally, this introduces a
certain error due to the diffusive transport of SGS turbulence energy.
In fact, the transport term in equation~(\ref{eq:sgs_energy_cl})
changes the unresolved energy without affecting the resolved energy
budget. We ignore this contribution in the energy update, if
backscattering is suppressed and the SGS model is not completely
coupled to the resolved flow. In the case of complete coupling,
however, the exact SGS transfer and dissipation rate are accounted for
in the dynamical equation for $e_{\mathrm{int}}$.  Of course, complete
coupling would seem appropriate regardless of the treatment of inverse
energy transfer.  However, if backscattering is suppressed, we found
that abandoning the SGS stresses and using the approximate energy
update as outlined above changes the results only little, while
increasing the computational efficiency significantly.

The test filter for the computation of the production parameter in the
semi-localised SGS model is numerically implemented as follows.
Orthogonal one-dimensional mesh filters with nine supporting nodes are
applied in the direction of each coordinate axis. The filter weights
are determined by matching the Fourier transform of the kernel as
accurately as possible to the spectral representation of an analytic
box filter. Then a single free parameters remains, which is the filter
scaling ratio $\gamma_{\mathrm{T}}$. For a given number of supporting
nodes, $N_{\mathrm{T}}$, an optimal value of $\gamma_{\mathrm{T}}$ can
be found from further constraining the Fourier transforms of the mesh
and the analytic box filter, respectively, to be equal at the cutoff
wavelength (appendix A.1.1 in \cite{Schmidt04}).  In the case
$N_{\mathrm{T}}=9$, we have $\Delta_{T}\approx
3.75\Delta_{\mathrm{eff}}\approx 6.74\Delta$.  Although
$\gamma_{\mathrm{T}}\approx 3.75$ is considerably larger than a factor
of two, which is commonly suggested in the literature, we obtained
optimal results with this setting rather than with test filters of
smaller characteristic length (section~4.3.3 in \cite{Schmidt04}).

\begin{figure}
  \begin{center}
    \vspace{110mm}
    \texttt{les\_stat3d\_flame\_speed.png}
    \bigskip
    \caption{ Evolution of dimensionless statistical moments for
      simulations with different flame speed models and SGS closures.
      The normalised RMS of the specific stirring force and
      the resolved momentum of the flow are shown in
      the top panels. The first three moments of the
      SGS turbulence velocity are plotted in the middle row of
      panels. Note that the mass-weighted mean of $q_{\mathrm{sgs}}$
      is scaled in units of the laminar burning speed
      $s_{\mathrm{lam}}$. In the bottom panels, the mean burning rate
      and the average mass fractions of fuel and processed nuclei are
      shown. }
    \label{fg:stat3d_flame_speed}
  \end{center}
\end{figure}

\begin{figure}
  \begin{center}
    \vspace{110mm}
    \texttt{les\_stat3d\_locl.png}
    \bigskip
    \caption{ Comparison between two different variants of the
      semi-localised SGS model. Backscattering is included in one
      case, whereas it is suppressed in the other case. The top panels show
      plots of the mean production parameter $C_{\nu}$ in regions
      containing ash, flames and fuel, respectively. The middle panels
      show the corresponding plots of the dissipation parameter, and
      the different contributions in equation~(\ref{eq:sgs_velocity})
      for the time evolution of $q_{\mathrm{sgs}}$ are plotted in the
      bottom panels. }
    \label{fg:stat3d_locl}
  \end{center}
\end{figure}

Statistical results from the simulations are shown in
figures~\ref{fg:stat3d_flame_speed} and~\ref{fg:stat3d_locl}. The
evolution of the mean burning rate $\langle B\rangle$ and the
corresponding fuel consumption together with the helium and nickel
production is shown in the bottom panels (g,h,i) of
figure~\ref{fg:stat3d_flame_speed}. For comparison, the RMS forcing,
momentum and Mach number are plotted in top row of
panels (a,b,c) for each simulation. In the course of the first
integral time, the slope of $\langle B\rangle$ in logarithmic scaling
is rather slowly rising. This indicates predominantly laminar
burning. The oscillations of $\langle B\rangle$ at early time are
caused by numerical discretisation errors, since the
burning regions initially tend to become elongated into thin shapes
and, in consequence, are only marginally resolved. As the flow becomes
increasingly turbulent and a growing fraction of the total flame
surface is subject to an enhanced propagation speed $s_{\mathrm{t}}\gg
s_{\mathrm{lam}}$, the rate of burning rises rapidly. Eventually, the
phase of exponentially growing energy release passes over into fading
combustion once the greater part of the fuel has been exhausted.  The
transition point between the quasi-laminar and the turbulent burning
phase can be estimated from the tangents to the almost linear portions
of the graph of $\langle B\rangle$ in logarithmic scaling. By means of
the plot of the mass-weighted statistical moments of
$q_{\mathrm{sgs}}$ in the panels (d,e,f), one can see that the
transition coincides with $\langle\rho
q_{\mathrm{sgs}}\rangle/(\rho_{0}s_{\mathrm{lam}})\approx 3$. The
somewhat larger threshold value of the mean SGS turbulence velocity
relative to the laminar burning speed for the onset of rapid turbulent
burning is possibly a consequence of intermittency. This is also
indicated by the large standard deviation, $\sigma(\rho
q_{\mathrm{sgs}})$, which is comparable to the mean, $\langle\rho
q_{\mathrm{sgs}}\rangle$, during the
production phase. The mass-weighted skewness, $\mathrm{skew}(\rho
q_{\mathrm{sgs}})$, is particularly large in the early phase when
eddies are forming locally, whereas it approaches a value near unity
in the regime of statistically stationary and homogeneous turbulence.

Comparing the left and the middle column of plots in
figure~\ref{fg:stat3d_flame_speed}, the evolution of the burning
process in the simulations which differ only by the
coupling of the SGS model to the momentum equation and the treatment
of inverse energy transfer appear quite similar. This can be seen in
figures~\ref{fg:les_nb_energy_evol} and \ref{fg:les_energy_evol} as
well, which show contour plots of the total energy per unit mass at
different instants of time for both simulations. However, the burning
process proceeds faster in the simulation without backscattering in
comparison to the simulation with the fully coupled SGS
model. Correspondingly, the peak of $\langle B\rangle$ is delayed in
the latter case.

\begin{figure}
  \begin{center}
    \vspace{110mm}
    \texttt{les\_nb\_energy\_evol.png}
    \bigskip
    \caption{Simulation of thermonuclear deflagration for an initial
      density $\rho_{0}\approx 2.90\cdot10^{8}\,\mathrm{g\,cm^{-3}}$.
      $V/s_{\mathrm{lam}}=100$, corresponding to a characteristic Mach
      number $V/c_{0}\approx 0.15$. The turbulent flame speed is given
      by $s_{\mathrm{t}}=
      \max(s_{\mathrm{lam}},q_{\mathrm{sgs}})$. The turbulent velocity
      $q_{\mathrm{sgs}}$ is computed with the semi-localised SGS
      model, where $C_{\nu}^{+}=\max(0,C_{\nu})$ is set as parameter
      of turbulence production. Shown are 2D contour sections of the
      normalised specific energy $\tilde{e}=e/c_{0}^{2}$ at different
      stages of the burning process. }
    \label{fg:les_nb_energy_evol}
  \end{center}
\end{figure}

\begin{figure}
  \begin{center}
    \vspace{110mm}
    \texttt{les\_nb\_q\_sgs\_evol.png}
    \bigskip
    \caption{  2D contour sections of SGS turbulence velocity relative
      to the laminar burning speed, $q_{\mathrm{sgs}}/s_{\mathrm{lam}}$, in
      logarithmic scaling. The panels correspond to those
      shown in Figure~\ref{fg:les_nb_energy_evol}.  }
    \label{fg:les_nb_q_sgs_evol}
  \end{center}
\end{figure}

The differences in the SGS model are illustrated in
figure~\ref{fg:stat3d_locl}.  The averages of the production parameter
$C_{\nu}$ in regions containing, respectively, ash, flames and fuel
are plotted in panels (a) and (b), respectively. If backscattering is
included, the mean of $C_{\nu}$ within the fuel is initially small. In
the course of turbulence production, the parameter is growing and,
eventually, $\langle C_{\nu}\rangle\approx 0.04$ in the statistically
stationary regime. On the other hand, if backscattering is suppressed,
the mean of $C_{\nu}$ changes only little in the dominating
topological region. The local values of $C_{\nu}$ are fluctuating
considerably for $\tilde{t}<1$. This can be seen from the strongly
oscillating averages of $C_{\nu}$ in ash and flames, which initially
fill quite narrow spatial regions and encounter varying conditions
while being advected by the flow.  The time evolution of the
dissipation parameters shown in panels (c) and (d), on the other hand,
exhibits more or less the same trend. During the first turn-over time,
$C_{\epsilon}$ vanishes identically. At time $\tilde{t}\approx 1$,
small turbulent vortices begin to form and turbulent dissipation sets
in. In statistical equilibrium, $C_{\epsilon}$ assumes a nearly
constant value of about 0.65 for the fully coupled model and 0.75
without backscattering. In the latter case, a larger dissipation rate
compensates the suppressed inverse energy transfer, as one can see
from the plots of the mean rate of production and dissipation
corresponding to the source terms on the right hand side of
equation~(\ref{eq:sgs_velocity}) in panels (e) and (f) of
figure~\ref{fg:stat3d_locl}. It is obvious that complete coupling
significantly reduces the rate of turbulence production. Nevertheless,
the mean value of $q_{\mathrm{sgs}}$ is found to be nearly the same in
the statistically stationary regime for both simulations.  As
mentioned in section~\ref{sc:sgs_locl}, we suspect that turbulence
production is systematically underestimated by the fully coupled SGS
model in combination with the PPM, because the kinetic energy injected
through backscattering into modes of high wave number will be quickly
dissipated by numerical viscosity. An impression of the spatiotemporal
evolution of $q_{\mathrm{sgs}}$ is given by the contour plots in
figures~\ref{fg:les_nb_q_sgs_evol} and \ref{fg:les_q_sgs_evol},
respectively.

\begin{figure}
  \begin{center}
    \vspace{110mm}
    \texttt{les\_energy\_evol.png}
    \bigskip
    \caption{ Simulation with the same parameter as specified in
    the caption of figure~\ref{fg:les_nb_energy_evol}, however,
    with $C_{\nu}$ as production parameter and complete coupling
    of SGS stresses to the flow. }
    \label{fg:les_energy_evol}
  \end{center}
\end{figure}

\begin{figure}
  \begin{center}
    \vspace{110mm}
    \texttt{les\_q\_sgs\_evol.png}
    \bigskip
    \caption{  2D contour sections of SGS turbulence velocity relative
      to the laminar burning speed, $q_{\mathrm{sgs}}/s_{\mathrm{lam}}$ in
      logarithmic scaling. The panels correspond to those
      shown in Figure~\ref{fg:les_energy_evol}.  }
    \label{fg:les_q_sgs_evol}
  \end{center}
\end{figure}

On account of results from laboratory measurements, Kim \textit{et
al.} argue that Pocheau's relation~(\ref{eq:sgs_flame_speed_pocheau})
with $n=2$ and $C_{\mathrm{t}}=20/3$ gives the most accurate prediction of
the turbulent flame speed \cite{KimMen99}.  The outcome
of running a simulation with the flame speed relation
\begin{equation}
  \label{eq:kim_men}
  s_{\mathrm{t}}=
  s_{\mathrm{lam}}\sqrt{1+\frac{20}{3}
  \left(\frac{q_{\mathrm{sgs}}}{s_{\mathrm{lam}}}\right)^{2}},
\end{equation}
is demonstrated by the statistics in the colum of panels (c,f,i) on
the very right of figure~\ref{fg:stat3d_flame_speed}. Backscattering
is also suppressed in this simulation. Now the peak of the burning
rate is reached even faster than in the case of the simulation with
$C_{\mathrm{t}}=1$.  In fact, the bulk of the burning takes place when
the level of SGS turbulence is quite low. Accordingly, the plots of
contour sections of the specific energy in
figure~\ref{fg:les_nb20_energy_evol} show that the flame surface is
smoother and less corrugated by the flow in the course of the burning
process. The slope of $\langle B\rangle$ is steepening
significantly just for $\langle\rho
q_{\mathrm{sgs}}\rangle/(\rho_{0}s_{\mathrm{lam}})\approx 1$.  This
would suggest that $C_{\mathrm{t}}=20/3$ is, indeed, a feasible
choice. On the other hand, $C_{\mathrm{t}}=4/3$ is favoured by Peters
\cite{Peters99}, which yields more or less the same behaviour as in
the case $C_{\mathrm{t}}=1$. From our current understanding, we should
consider $C_{\mathrm{t}}$ as a parameter of the flame speed model,
which is to be chosen within reasonable limits and validated \emph{a
posteriori} by the results obtained for a particular application.

\begin{figure}
  \begin{center}
    \vspace{110mm}
    \texttt{les\_nb20\_energy\_evol.png}
    \bigskip
    \caption{The physical parameters are the same as for the
    simulations in the previous figures, but this time the flame speed
    relation is given by
    $s_{\mathrm{t}}=s_{\mathrm{lam}}[1+\frac{20}{3}(q_{\mathrm{sgs}}/s_{\mathrm{lam}})^{2}]^{1/2}$. Shown
    are 2D contour sections of the normalised specific energy
    $\tilde{e}=e/c_{0}^{2}$ at different stages of the burning
    process.  }
    \label{fg:les_nb20_energy_evol}
  \end{center}
\end{figure}

\section{Conclusion}

The numerical simulation of thermonuclear deflagration in a box
subject to stochastic stirring was utilised as a test problem for the
study of flame speed models. The evolution of the flame front was
computed by means of the level set method in the so-called passive
implementation.  Essentially, an effective flame propagation speed
must be calculated, if the Gibson scale is small compared to the
resolution of the computational grid. A subgrid scale (SGS) model
based on the budget of turbulence energy determines a velocity scale
which is proportional to the propagation speed of flame fronts in the
fully turbulent regime. Some of the closure parameters of the SGS
model are locally calculated with dynamical procedures. Thus, we have
a semi-localised model.

Particularly, we compared two variants of this model.  In one case,
inverse energy transfer from subgrid toward resolved scales was
included, in the other case it was suppressed. Inverse energy transfer
is also known as backscattering.  For a consistent treatment of
backscattering, complete coupling of the SGS model and the resolved
hydrodynamics is indispensable. In combination with the piece-wise
parabolic method, this entails difficulties stemming from the
significant numerical viscosity of the scheme. But we obtained
sensible results when suppressing backscattering and applying a
simplified SGS model with partial coupling.

Depending on the constant of proportionality $\sqrt{C_{\mathrm{t}}}$
in the asymptotic flame speed relation, we found the transition from
laminar to turbulent burning at noticeably different points in the
course of turbulence production. This transition comes about once the
SGS turbulence velocity must exceed the laminar burning speed in a
significant volume fraction of the computational domain. Then the
nuclear energy generation grows at a much higher rate, and the flame
surface develops an intricate structure due to the stretching and
folding caused by turbulent vortices. If $C_{\mathrm{t}}$ is about
unity, the peak of nuclear energy release appears roughly when the
turbulent flow becomes statistically stationary and homogeneous. For
larger values of $C_{\mathrm{t}}$, most of the fuel is consumed in
advance of turbulence becoming fully developed. We propose to consider
$C_{\mathrm{t}}$ as a control parameter, which regulates the overall
rapidness of the burning process.

The semi-localised SGS model presented here is especially suitable for
any kind of transient and inhomogeneous turbulent combustion
process. It is the first implementation of this kind of SGS model for
an astrophysical application, namely, the numerical simulation of
thermonuclear supernovae.

\section{Acknowledgements}

The simulations were run on the Hitachi SR-8000 of the \emph{Leibniz
Computing Centre} the IBM p690 of the \emph{Computing Centre of the
Max-Planck-Society} in Garching, Germany. We thank M. Reinecke for his
helpful remarks concerning the level set method.  One of the authors
(W Schmidt) was supported by the priority research program
\emph{Analysis and Numerics for Conservation Laws} of the Deutsche
Forschungsgesellschaft. The research of W Schmidt and J C Niemeyer was
supported by the Alfried Krupp Prize for Young University Teachers of
the \emph{Alfried Krupp von Bohlen und Halbach Foundation}.

\end{document}